\documentstyle[prl,aps,manuscript,epsf]{revtex}

\begin{document}
\title{Q-Deformed Anisotropic Superexchange Interaction, Frustration and GL$%
_{pq}\left( 2\right) $}
\author{S. Grillo and H. Montani}
\address{{\it Centro At\'{o}mico Bariloche and Instituto Balseiro,}\\
8400-S. C. de Bariloche, Argentina}
\date{April 1998 }
\maketitle

\begin{abstract}
We study a suitable q-deformed version of the Moriya's superexchange
interaction theory by means of its underlying quantum group structure. We
show that the one-dimensional chain case is associated with the non-standard
quantum group $GL_{pq}\left( 2\right) $, evidencing the integrability
structure of the system. This biparametric deformation of $GL\left( 2,{\bf C}
\right) $ arise as a twisting of $GL_{q}\left( 2\right) $ and it match
exactly the local rotation appearing in the Shekhtman's work \cite{Sh}. This
allow us to express the frustration condition in terms of this twisting,
also showing that effect of the Moriya's vector amounts to a twisting of the
boundary condition.
\end{abstract}



The presence of weak ferromagnetism in various antiferromagnetic compounds
was explained at the end of 50's by Dzyaloshynskii \cite{Dz}, from purely
symmetry grounds, introducing in the thermodynamic potential of those
systems an antisymmetric spin-spin interaction term ${\bf D}\cdot \left(
{\bf M}_{1}\times {\bf M}_{2}\right) $, were ${\bf M}_{1}$ and ${\bf M}_{2}$
are sublattices magnetization, and ${\bf D}$ a macroscopic parameter called
Dzyaloshynskii's vector. The microscopic basis for the Dzyaloshynskii
conjecture was given by Moriya \cite{Mo}. He extended the Anderson's
superexchange interaction theory \cite{An} in order to include in the
one-electron Hamiltonian an spin-orbit coupling. In terms of electron
annihilators and creations operators, $c\;$and $c^{\dagger }$, the resulting
Hamiltonian is (considering here just one orbital state per ion, denoted in
\cite{Mo} by $n$) $H=H_{o}+H_{t}$, being:
\begin{eqnarray*}
H_{o} &=&\sum_{r}\sum_{\alpha }\varepsilon _{r}\;c_{r\alpha }^{\dagger
}\;c_{r\alpha } \\
H_{t} &=&\sum_{r,r^{\prime }}\sum_{\alpha ,\alpha ^{\prime }}t_{rr^{\prime
}}\;c_{r\alpha }^{\dagger }\;\left( e^{i\theta _{rr^{\prime }}\;\widehat{%
{\bf d}}_{rr^{\prime }}{\bf \cdot \sigma }}\right) _{\alpha ,\alpha ^{\prime
}}\;c_{r^{\prime }\alpha ^{\prime }}
\end{eqnarray*}
where $r,r^{\prime }$ runs over lattice sites, ${\bf \sigma =}\left( \sigma
_{x},\sigma _{y},\sigma _{z}\right) $ are the Pauli matrices (so $\alpha
,\alpha ^{\prime }=\left\{ \uparrow ,\downarrow \right\} $), $\varepsilon
_{r}$ are crystal energies, and $t_{rr^{\prime }},\theta _{rr^{\prime }},%
\widehat{{\bf d}}_{rr^{\prime }}$ come from the transfer integrals $%
b_{rr^{\prime }}=b_{r^{\prime }r}^{*}$ and ${\bf C}_{rr^{\prime }}={\bf C}%
_{r^{\prime }r}^{*}$ \cite{Mo}, which in the case of {\em non degenerate
orbital ground state} (except for being a Kramer's doublet) $b_{rr^{\prime }}
$ is real and ${\bf C}_{rr^{\prime }}$ is purely imaginary and we can write $%
b_{rr^{\prime }}=t_{rr^{\prime }}\cos \theta _{rr^{\prime }}$ and ${\bf C}%
_{rr^{\prime }}=i\;t_{rr^{\prime }}\sin $ $\theta _{rr^{\prime }}\;\widehat{%
{\bf d}}_{rr^{\prime }}$ with $t_{rr^{\prime }},\theta _{rr^{\prime }}\in
{\bf R}$ and $\widehat{{\bf d}}_{r^{\prime }r}\in {\bf R}^{3}$ ($\widehat{%
{\bf d}}_{r^{\prime }r}\cdot \widehat{{\bf d}}_{r^{\prime }r}=1$). The $%
\theta _{rr^{\prime }}=0$ case corresponds to the Anderson's theory, while $%
\theta _{rr^{\prime }}\neq 0$ give rise to an anisotropy term as consequence
of the spin-orbit interaction. From now on we omit, for simplicity, the $r$%
's indices in all parameters. Following \cite{Mo} and \cite{An}, and taking
into account the correct factor $4$ pointed out in \cite{Sh}, we arrive to
an effective Hamiltonian, up to a constant term:
\begin{eqnarray*}
{\cal H}=\sum_{\left\langle r,r^{\prime }\right\rangle }\frac{t^{2}}{U}\{
&&cos2\theta \;\left( {\bf \sigma }_{r}\cdot {\bf \sigma }_{r^{\prime }}-%
\widehat{{\bf d}}{\bf \cdot \sigma }_{r}\;\widehat{{\bf d}}{\bf \cdot \sigma
}_{r^{\prime }}\right) + \\
&&+{\bf \widehat{d}\cdot \sigma }_{r}\;{\bf \widehat{d}\cdot \sigma }%
_{r^{\prime }}{\bf +I+}sin2\theta \;\widehat{{\bf d}}{\bf \cdot }\left( {\bf %
\sigma }_{r}\times {\bf \sigma }_{r^{\prime }}\right) \;\}
\end{eqnarray*}
where $\left\langle r,r^{\prime }\right\rangle $ runs now over nearest
neighbor, ${\bf \sigma }_{r}$ are Pauli matrices acting on the $r$-th site
\footnote{%
In general, given a collection of isomorphics linear spaces $\left\{
V_{i}\right\} $ and an operator ${\cal O}$ that acts on $\otimes
_{i=1}^{M}V_{i}$, we will denote$\;{\cal O}_{i_{1}...i_{M}}$ as the operator
that acts non trivially on $\otimes _{k=1}^{M}V_{i_{k}}$ and as the identity
in the rest.} , ${\bf I}$ the identity operator and $U$ is the so-called
Hubbard energy (to put two electrons on the same ion). The Hamiltonian $%
{\cal H}$ represents the {\em anisotropic superexchange interaction} ($ASI$%
), where the last term is called Dzyaloshynskii-Moriya (DM) interaction (the
microscopic counterpart of the Dzyaloshynskii interaction), and $%
t^{2}/U\;sin2\theta \;\widehat{{\bf d}}$ is the Moriya's vector for the bond
$r,r^{\prime }$ \footnote{%
This antisymmetric coupling has been observed ten years ago in the
high-temperature superconducting material La$_{2}$CuO$_{4}$ \cite{tio}, and
was studied on some cuprates in \cite{Ri}. Then, this mecanism was used to
descript several properties in doped La$_{2}$CuO$_{4}$ as La$_{2-x}$Ba$_{x}$
CuO$_{4}$ \cite{Rice}. Recentlly Affleck {\em et al}. \cite{Aff} have used
the DM interaction to explain the field-induced gap in antiferromagnetic
(quasi-one-dimensional) Cu Benzoato chains.}. For $\theta =0$, ${\cal H}%
=H_{xxx}$, i.e., the $XXX$ or isotropic Heisenberg model.

At this point Shekhtman ${\em et}$ {\em al. }\cite{Sh} showed that, term by
term, ${\cal H}$ can be cast in an isotropic Heisenberg model form, i.e.:

\[
{\cal H}^{r,r^{\prime }}=\frac{t^{2}}{U}\;\left\{ {\bf \sigma }_{r}\cdot
{\bf \sigma }_{r^{\prime }}^{\prime }+{\bf I}\right\}
\]
with ${\bf \sigma }_{r^{\prime }}^{\prime }=e^{-i\theta \;\widehat{{\bf d}}
\cdot {\bf \sigma }_{r^{\prime }}}{\bf \sigma }_{r^{\prime }}\;e^{i\theta \;
\widehat{{\bf d}}\cdot {\bf \sigma }_{r^{\prime }}}$, i.e., ${\bf \sigma }
_{r^{\prime }}$ rotated in $-2\theta $ \footnote{%
Actually, in Shekhtman's work it makes a rotation in $\theta $ for $r$ and $%
-\theta $ for $r^{\prime }$. It can see the relevant is the relative angle
between the rotations. This angle is just $-2\theta $.}. This local
rotations, defined bond to bond, are similarity transformations term by
term, i.e. ${\cal H}^{r,r^{\prime }}=e^{-i\theta \;\widehat{{\bf d}}\cdot
{\bf \sigma }_{r^{\prime }}}\;H_{xxx}^{r,r^{\prime }}\;e^{i\theta \;\widehat{%
{\bf d}}\cdot {\bf \sigma }_{r^{\prime }}}$, but not in general for the
whole Hamiltonians. A sufficient condition to extend the local rotations to
a similarity transformation on the whole lattice is that the product of such
rotations along any closed path on the lattice be equal to $\left( -1\right)
^{n}{\bf I}$ ($n\in {\bf Z}$) (this condition is just a compatibility
requirement of the local transformations to make this extension). In the 1-D
open chains this condition always holds (there is no closed paths), but for
closed ones it depends on the boundary. If this condition holds it means
that there is {\em no frustration }(or the transformation is {\em non
frustrated}), and the Hamiltonians are similar, hence ${\cal H}$ and $%
H_{xxx} $ have the same eigenspectra, being the similarity transformation in
1-D closed chain
\begin{equation}
{\bf A}=\prod_{r=2}^{N}\prod_{j=1}^{r-1}e^{-i\theta _{r-j}\;\widehat{{\bf d}}
_{r-j}{\bf \cdot \sigma }_{r}}  \label{sim}
\end{equation}
with $N$ the number of sites, and denoting $\tau _{r,r^{\prime }}=\tau
_{r,r+1}\equiv \tau _{r}$ for all parameters $\tau $. So they conclude that
the {\em frustration is a necessary condition to have weak ferromagnetism }
(because the $XXX$'s ground state have zero magnetization -is purely
antiferromagnetic-).

The main aim of our work is to show up the quantum group structure
associated to this model and study the frustration conditions in terms of
its algebraic content. To this end we are going to make a suitable
generalization of the model allowing us to reveal these underlying algebraic
structure, which is deeply tied up to the integrability of the model. We
generalize $H$ by making an analytic continuation through $\theta
\rightarrow \theta _{c}$, such that writing $\theta _{c}=\theta +i\;\phi $ ($%
\alpha ,\beta \in {\bf R}$) and defining $\;$
\begin{equation}
\begin{array}{c}
p=e^{2i\theta } \\
q=e^{-2\phi }
\end{array}
\label{def}
\end{equation}
$H_{t}$ becomes in:
\[
H_{t}=\sum_{\left\langle r,r^{\prime }\right\rangle }\sum_{\alpha ,\alpha
^{\prime }}\,t\;c_{r\alpha }^{\dagger }\;\left( \;p^{\widehat{{\bf d}}{\bf %
\cdot \sigma }/2}q^{\widehat{{\bf d}}{\bf \cdot \sigma }/2}\right) _{\alpha
,\alpha ^{\prime }}\;c_{r^{\prime }\alpha ^{\prime }}
\]
and ${\cal H}$, up to a constant term:
\begin{eqnarray}
{\cal H} &&_{q}=\sum_{\left\langle r,r^{\prime }\right\rangle }\frac{t^{2}}{%
U }\,\{\frac{p+p^{-1}}{2}\left( {\bf \sigma }_{r}\cdot {\bf \sigma }
_{r^{\prime }}-\widehat{{\bf d}}{\bf \cdot \sigma }_{r}\widehat{{\bf d}}{\bf %
\cdot \sigma }_{r^{\prime }}\right)  \nonumber \\
&&+\frac{q+q^{-1}}{2}\left( \widehat{{\bf d}}{\bf \cdot \sigma }_{r}\widehat{%
{\bf d}}{\bf \cdot \sigma }_{r^{\prime }}+{\bf I}\right) {\bf +}\frac{
p-p^{-1}}{2i}\widehat{{\bf d}}{\bf \cdot }\left( {\bf \sigma }_{r}\times
{\bf \sigma }_{r^{\prime }}\right) \}  \label{q-de}
\end{eqnarray}
and we name it $q$-deformed $ASI\,\,(q-ASI)$. We can see that $%
p=1\Rightarrow {\cal H}_{q}=H_{xxz}$, i.e., the $XXZ$ or anisotropic
Heisenberg model (we actually have an inhomogeneous version of the $XXZ$,
unless $\widehat{{\bf d}}\ $and $q$ were the same for all bonds, so we can
make an appropriate global rotation to pass from $\widehat{{\bf d}}$ to $%
\widehat{{\bf z}}$ and have the proper $XXZ$).

Let us analyze the effects of the frustration for the $q$-deformed
Hamiltonian. As for $q=1$, we can write term by term ${\cal H}
_{q}^{r,r^{\prime }}=p^{-\widehat{{\bf d}}{\bf \cdot \sigma }_{r^{\prime
}}/2}\;H_{xxz}^{r,r^{\prime }}\;p^{\widehat{{\bf d}}{\bf \cdot \sigma }
_{r^{\prime }}/2}$. In this case, the non frustration condition is not
enough to extend this rotations to a similarity transformation between $%
{\cal H}_{q}$ and $H_{xxz}$, the Moriya's directions $\left\{ \widehat{{\bf %
d }}_{r,r^{\prime }}\right\} $ being all equals is needed, i.e., $\widehat{%
{\bf d}}_{r,r^{\prime }}=\widehat{{\bf d}}$ $\forall r,r^{\prime }$ (the
nature of this fact is the $SU\left( 2\right) $-invariance of $H_{xxx}$,
while $H_{xxz}$ is just $U\left( 1\right) \otimes {\bf Z}_{2}$-invariant).
In the 1-D case, the similarity transformation is given by putting in ${\bf %
A }$ (see ec. (\ref{sim})) $\widehat{{\bf d}}_{r}=\widehat{{\bf d}}$ $%
\forall r $. Then, for the $q$-$ASI$ model, when $\widehat{{\bf d}}%
_{r,r^{\prime }}= \widehat{{\bf d}}$ $\forall r,r^{\prime }$, the previous
conclusion still holds, because if there is no frustration ${\cal H}_{q}$\
is similar to $H_{xxz}$\ which have a purely antiferromagnetic ground state
and there is no net ferromagnetic moment (remember $q=e^{-2\phi }$).

Let's briefly review how to construct Hamiltonian integrable models from a
given ${\bf R}\left( x\right) $ solution of the quantum Yang-Baxter equation
(QYBE)
\[
{\bf R}_{12}\left( x/y\right) {\bf R}_{13}\left( x\right) {\bf R}_{23}\left(
y\right) ={\bf R}_{23}\left( y\right) {\bf R}_{13}\left( x\right) {\bf R}
_{12}\left( x/y\right)
\]
with ${\bf R}\left( x\right) $ acting on $V_{a}\otimes V$, $V_{a}$ and $V$
being isomorphic linear spaces. Regarding this ${\bf R}\left( x\right) $
-matrix as a representation of the Lax operators of a quantum spin chain or
as the Boltzman weights of some statistical model \cite{fad} \cite{korepin}
, one built up the monodromy matrix (see also \cite{marti} and references
therein) ${\bf T}_{a}^{{}}(x)={\bf R}_{a,1}\left( x\right) \cdot {\bf R}
_{a,2}\left( x\right) \cdot \cdot \cdot {\bf R}_{a,N}\left( x\right) $
giving rise to transfer matrix:

\[
{\bf t}\left( x\right) =tr_{a}\left[ {\bf T}_{a}(x)\right]
\]
where the trace is over the auxiliary $V_{a}$ space tied to each site, so $%
{\bf t}^{(N)}$ is an operator acting on $V^{\otimes ^{N}}$. These monodromy
matrix can be encoded into a bialgebra structure, the Yang-Baxter algebra
(YBA), which in some limit of the spectral parameter becomes quasitriangular
{\footnotesize \ }\cite{frt}{\footnotesize . }In this framework, the
transfer matrices $\{{\bf t}\left( x\right) \}$, with different spectral
parameter $x$, appears as a set of commuting quantities from which one
derive an associated Hamiltonian for the system
\[
{\sf H}=c\frac{d}{dx}ln\left[ {\bf t}\left( x\right) \right] _{x=x_{o}}
\]
thus leading to the integrability of the system \cite{korepin}( $x_{o}$ is
some appropriate value of $x$ and $c\in {\bf C}$). The locality of the
Hamiltonian is warranted by choosing (if there exist) $x_{o}{\bf \ }$such
that ${\bf R}\left( x_{o}\right) $ $=\alpha {\bf P}$, with ${\bf P}$ the
permutation matrix, then ${\sf H}=\sum_{k=1}^{N}H_{k,k+1}$ with $%
H_{k,k+1}=c/\alpha \frac{d}{dx}{\bf R}_{k,k+1}\left( x_{o}\right) {\bf P}
_{k,k+1}$ acting non trivially only in spaces $k$-th and $\left( k+1\right) $
-th of $V^{\otimes ^{N}}$, and $H_{N,N+1}\equiv H_{N,1}$. For later
convenience, we introduce in ${\bf R}_{a,k}\left( x\right) $ a
transformation which preserves integrability ${\bf R}_{a,k}{\bf \rightarrow
\Gamma }_{k}^{(k)}{\bf R}_{a,k}\left( {\bf \Gamma }_{k}^{(k)}\right) ^{-1}$,
and defining ${\bf S=}\prod_{k=1}^{N}{\bf \Gamma }_{k}^{(k)}$ it drives to a
Hamiltonian ${\sf H}^{{\bf S}}={\bf S}\cdot \sum_{k=1}^{N}H_{k,k+1}\cdot
{\bf S}^{-1}$.

It is well known that the quantum deformation of $GL(2,{\bf C)}${\bf , }
namely $GL_{q}\left( 2\right) $, is the underlying algebraic structure of
the $XXZ$ model (see \cite{fad}, and references therein), with associated $R$
-matrix:
\[
{\bf R}^{q}=\left[
\begin{array}{cccc}
q & \cdot & \cdot & \cdot \\
\cdot & 1 & q-q^{-1} & \cdot \\
\cdot & \cdot & 1 & \cdot \\
\cdot & \cdot & \cdot & q
\end{array}
\right]
\]
In fact, defining ${\bf R}^{q}\left( x\right) =x\;{\bf R}^{q}-\frac{1}{x}\;
{\bf P}\cdot \left( {\bf R}^{q}\right) ^{-1}\cdot {\bf P}$ (with its
associated YBA), we built up, for $x=1$, the homogeneous Hamiltonian ${\sf H}
_{q}^{{\bf S}}={\bf S}\cdot \sum_{k=1}^{N}H_{k,k+1}^{q}\cdot {\bf S}^{-1}$,
or explicitly

\begin{eqnarray*}
{\sf H}_{q}^{{\bf S}}=\frac{c}{q-q^{-1}}\sum_{k=1}^{N}\{&&{\bf \sigma }
_{k}\cdot {\bf \sigma }_{k+1}-\;\widehat{{\bf d}}{\bf \cdot \sigma }_{k}
\widehat{{\bf d}}{\bf \cdot \sigma }_{k+1}  \nonumber \\
&&+\frac{q+q^{-1}}{2}\,\left( \widehat{{\bf d}}{\bf \cdot \sigma }_{k}%
\widehat{{\bf d}}{\bf \cdot \sigma } _{k+1}+{\bf I}\right) \}
\end{eqnarray*}
where$\;$if $\widehat{{\bf d}}=(cos\gamma \;sin\varphi ,sin\gamma
\;sin\varphi ,cos\varphi )$ \footnote{%
It's totally equivalent to work with $\widehat{{\bf z}}$ however, in order
to preserve the connection with the Moriya's vector concept, we prefer take
an arbitrary direction.} so ${\bf S=}exp\left[ i\varphi \;\widehat{{\bf u}}
\cdot {\sum_{r=1}^{N}}{\bf \sigma }_{r}/2\right] $ and $\widehat{{\bf u}}
=(-sin\gamma ,cos\gamma )$. From ec. (\ref{q-de}) for $p=1$, it sees that $%
{\sf H}_{q}^{{\bf S}}=H_{xxz}$ if $c=t^{2}/U\ (q-q^{-1})$.

The question naturally arise what is the underlying algebraic structure of
the $q$-$ASI$ ? Introducing a 2-cocycle twisting transformation $\Phi $ on $%
GL_{q}\left( 2\right) $ \cite{drinf} \cite{resh}, which maps the algebraic
structure into another equivalent now characterized by the $R$-matrix ${\bf %
R }^{pq}$, given by:
\[
{\bf R}^{q}\rightarrow {\bf R}^{pq}=\widetilde{\Phi }{\bf R}^{q}\Phi ^{-1}
\]
being $\Phi =\rho \otimes {\bf I}$ with $\rho =p^{-\widehat{{\bf z}}{\bf %
\cdot \sigma }/2}$, $\;\widetilde{\Phi }=\Phi _{21}={\bf P\cdot }\Phi \cdot
{\bf P=I}\otimes \rho $ ; which is associated to the non-standard quantum
group $GL_{pq}\left( 2\right) $ \cite{demindov}. Proceeding as in ${\bf R}
^{q}$ case described above, we introduce ${\bf R}^{pq}\left( x\right) =x\;
{\bf R}^{pq}-\frac{1}{x}\;{\bf P}\cdot \left( {\bf R}^{pq}\right) ^{-1}\cdot
{\bf P}$ and build up the homogeneous Hamiltonian ${\sf H}_{pq}^{{\bf S}}$:
\begin{eqnarray}
{\sf H}_{pq}^{{\bf S}}=\frac{c}{q-q^{-1}}{\sum_{k=1}^{N}}\{ &&\frac{p+p^{-1}
}{2}\left( {\bf \sigma }_{k}\cdot {\bf \sigma }_{k+1}-\widehat{{\bf d}}{\bf %
\;\cdot \sigma }_{k}\widehat{{\bf d}}{\bf \cdot \sigma }_{k+1}\right) +
\nonumber \\
&&+\frac{q+q^{-1}}{2}\left( \widehat{{\bf d}}{\bf \cdot \sigma }_{k}\;
\widehat{{\bf d}}{\bf \cdot \sigma }_{k+1}+{\bf I}\right)  \nonumber \\
&&+\frac{p-p^{-1}}{2i}\widehat{{\bf d}}{\bf \cdot }\left( {\bf \sigma }
_{k}\times {\bf \sigma }_{k+1}\right) \}  \label{hom}
\end{eqnarray}
$\;$It is worth remarking that, since the construction, the monodromy matrix
${\bf T}(x)$ of this system satisfy a quadratic algebra relation (YBA),
\[
{\bf R}^{pq}\left( x/y\right) {\bf T}(x)\otimes {\bf T}(y)={\bf T}(y)\otimes
{\bf T}(x){\bf R}^{pq}\left( x/y\right)
\]
from which one derives the integrability of the system.

In this way, we found an integrable spin chain associated to the quantum
group $GL_{pq}\left( 2\right) $ that strongly resembles ${\cal H}_{q}$. In
fact, in the particular case where $p$ and $q$ are those of (\ref{def}) and $%
c=t^{2}/U\ (q-q^{-1})$, ${\sf H}_{pq}^{{\bf S}}={\cal H}_{q}$ (see ec. (\ref
{q-de})) for the homogeneous periodic chain case. This tell us that the
quantum group $GL_{pq}\left( 2\right) $ is the underlying algebraic
structure of the $q$-$ASI$ model, allowing us to understood it as twisted
version of $XXZ$ model. There is a nice connection of the twisting $\Phi $
with the map found in \cite{Sh} as we shall explain below.

Let's work out the effect of the twisting on the Hamiltonian. Building up a
Hamiltonian ${\sf H}^{{\bf S}}={\sum_{k=1}^{N}}H_{k,k+1}^{{\bf S}}={\bf S}%
\cdot {\sum_{k=1}^{N}}H_{k,k+1}\cdot {\bf S}^{-1}$ , from an ${\bf R}$
-matrix such that ${\bf R}\left( x=1\right) \propto $ ${\bf P}$, then from $%
\widetilde{\Phi }{\bf R}\Phi ^{-1}$ the Hamiltonian is
\begin{equation}
{\sf H}_{\Phi }^{{\bf S}}={\sum_{k=1}^{N}}\widetilde{\Phi }_{k,k+1}^{{\bf S}%
}\cdot H_{k,k+1}^{{\bf S}}\cdot \left( \widetilde{\Phi }^{{\bf S}}\right)
_{k,k+1}^{-1}  \label{tw}
\end{equation}
In our case ${\sf H}^{{\bf S}}=H_{xxz}$, ${\sf H}_{\Phi }^{{\bf S}}={\cal H}%
_{q}$ and $\widetilde{\Phi }_{k,k+1}^{{\bf S}}={\bf S}\cdot \widetilde{\Phi }%
_{k,k+1}\cdot {\bf S}^{-1}=p^{-\widehat{{\bf d}}{\bf \cdot \sigma }_{k+1}/2}$
. So, the consequences of the twist on the Hamiltonian boils down exactly to
the rotation pointed out in Shekhtman's work \cite{Sh}, and the necessary
and sufficient condition to avoid frustration is that
\[
p=e^{i2n\pi /N}\,\,,\,n\in {\bf Z\,\,}
\]
In other words, if $p=e^{i2n\pi /N}$ ($n\in {\bf Z}$ and $\,N\in {\bf N}$),
then there is no frustration if and only if we built an $N$-sites
Hamiltonian. Then, although the twisting leads to integrable spin chains for
arbitrary values of the parameter $p$, the {\it no frustration} constraint
requires $p$ being a root of the unit and it is just in this case when the
twisting becomes in a similarity transformation between $H_{xxz}$ and ${\cal %
H}_{q}$.

As we could guess for 1-D closed chains, the relevant information about the
effect of the Moriya's vector can be put it in terms of boundary conditions.
In fact, noting that $\left[ H_{k,k+1},\rho _{k}\;\rho _{k+1}\right] =0$
(remember $\widetilde{\Phi }_{k,k+1}=\rho _{k+1}$), we can write ${\bf S=}
\widehat{{\bf S}}\cdot {\bf X}$ with ${\bf X=\prod }_{l=2}^{N}(\rho
_{l})^{l-1}$, taking ${\sf H}_{\Phi }^{{\bf S}}$ the form:

\begin{equation}
{\sf H}_{\Phi }^{{\bf S}}=\widehat{{\bf S}}\cdot \left[ {\sum_{k=1}^{N}}
H_{k,k+1}+\left( \Omega _{1}\right) ^{-1}H_{N,1}\Omega _{1}\right] \cdot
\widehat{{\bf S}}^{-1}  \label{bc}
\end{equation}
with $\Omega =\rho ^{N}$ . So ${\cal H}_{q}$ is always similar to $%
H_{xxz}^{tbc}$, i.e., the $XXZ$ with twisted boundary conditions ($tbc$)
given by $\Omega $. In this terms, i.e. as an $XXZ$ model with $tbc$, the
Hamiltonian ${\sf H}_{\Phi }^{{\bf S}}$ have just been widely studied by
means of the Bethe equations \cite{Alc} \cite{Pas}. Again, when there is no
frustration, i.e. $\,\Omega =\left( -1\right) ^{n}{\bf I}${\bf , }one
recovers a similarity with the standard $H_{xxz}$.

Most of the systems what manifest a DM interaction have $p$'s that change
bond to bond, like the so-called canonical DM antiferromagnet that alternate
$p$ and $1/p$ on successive bonds. So to have Hamiltonians that describe
this systems we shall introduce inhomogeneities in ${\sf H}_{pq}^{{\bf S}}$.
It is possible to introduce some kind of inhomogeneity, preserving
integrability, by means the transformation ${\bf R}\left( x\right)
\rightarrow \widehat{{\bf R}}^{(k)}\left( x\right) ={\bf R}\left( x\right)
\left( \Psi ^{\left( k\right) }\otimes {\bf I}\right) $, where $\Psi
^{\left( k\right) }\in {\cal G}_{{\bf R}}$, being ${\cal G}_{{\bf R}
}=\left\{ \Psi \right\} $ the group of operators satisfying $\left[ {\bf R}
\left( x\right) ,\Psi \otimes \Psi \right] =0$ $\forall x$, and representing
the {\em internal symmetries} of the YBA associated to ${\bf R}\left(
x\right) $ (see \cite{DV} and references therein). The Hamiltonian we built
with $\widehat{{\bf R}}$'s is

\begin{equation}
\widehat{{\sf H}}^{{\bf S}}={\bf S}\cdot \sum_{k=1}^{N}\left( \Psi
_{k+1}^{\left( k\right) }\right) ^{-1}H_{k,k+1}\;\Psi _{k+1}^{\left(
k\right) }\cdot {\bf S}^{-1}  \label{inh}
\end{equation}
How in $\widehat{{\sf H}}^{{\bf S}}$ appear $\Psi $ and $\Psi ^{-1}$ the
group we can consider is the quotient $\widehat{{\cal G}}_{{\bf R}}={\cal G}
_{{\bf R}}/{\cal Z}$, where ${\cal Z}$ is the group of scalar matrices on $%
V_{a}$. It is straightforward to show that for ${\bf R}^{pq}$:

\[
\widehat{{\cal G}}_{{\bf R}^{pq}}=\left\{
\begin{array}{l}
\{\mu ^{-\widehat{{\bf z}}{\bf \cdot \sigma }/2}\}\,if\,p\neq 1 \\
\{\mu ^{-\widehat{{\bf z}}{\bf \cdot \sigma }/2}\}\cup \{\xi ^{-\widehat{%
{\bf x}}{\bf \cdot \sigma }/2}\}\,if\,p=1
\end{array}
\right.
\]
$\mu ,\xi \in {\bf C}-\{0\}$, and calling $\left[ \widehat{{\cal G}}_{{\bf R}
^{pq}}\right] _{p\neq 1}=\left\{ \Psi ^{\mu }\right\} _{\mu \in {\bf C}}$,
from ecs. (\ref{tw}) and (\ref{inh}) we see that the change in ${\sf H}
_{pq}^{{\bf S}}$ is equivalent to put $p\mu _{k}$ instead of $p$ in (\ref
{hom}). In such a case there is frustration $iff$
\[
\prod_{k=1}^{N}[p\mu _{k}]\neq e^{i2n\pi /N}
\]
$\;\forall n\in {\bf Z}$. Choosing $\mu _{2k}=1$ and $\mu _{2k+1}=p^{-2}$ we
have the canonical case \footnote{%
In such a case there is no frustration for all $p$ $iff$ $N$ is even.}.

{\bf \ }Because of $\left[ {\bf R}\left( x\right) ,\Psi \otimes \Psi \right]
=0$ $\forall x$ implies that $\left[ H_{k,k+1},\Psi _{k}\;\Psi _{k+1}\right]
=0$, the $\Psi $'s can be cumulated in the last term of $\widehat{{\sf H}}^{%
{\bf S}}$, as above (ec. (\ref{bc})), through ${\bf X=\prod }
_{l=2}^{N}\prod_{k=1}^{l-1}\Psi _{l}^{\left( l-k\right) }$ , having $\Omega
=\prod_{k=1}^{N}\Psi ^{\left( N+1-k\right) }$. So the associated Hamiltonian
is, up to a similarity transformation, the original one with twisted
boundary conditions. In our case, the twisted boundary conditions will be
given by $\Omega =\prod_{k=1}^{N}[p\mu _{k}]^{-\widehat{{\bf z}}{\bf \cdot
\sigma }/2}$. Going back, it sees that it can associate this Hamiltonian to $%
GL_{p^{\prime }q}\left( 2\right) $, with $p^{\prime }$ any $N$-root $\left\{
\prod_{k=1}^{N}[p\mu _{k}]\right\} ^{1/N}$ , instead of $GL_{pq}\left(
2\right) $.

When $q=1$, the Hamiltonian $\left[ {\sf H}_{pq}^{{\bf S}}\right]
_{q=1}\equiv $ ${\sf H}_{p}^{{\bf S}}$ can be made completely inhomogeneous
by using again $\left\{ \Psi ^{\mu }\right\} _{\mu \in {\bf C}}$ and a
similarity transformation ${\bf Y=\prod }_{k=2}^{N}\Lambda _{k}^{(k)}$ with
\[
\Lambda _{k}^{(k)}=\left[ \prod_{j=1}^{k-1}[p\mu _{k-j}]^{-\widehat{{\bf d}}
_{k-j}{\bf \cdot \sigma }_{k}/2}\right] \cdot \left\{ \prod_{j=1}^{N}[p\mu
_{j}]\right\} ^{\left( k-1\right) \;\widehat{{\bf d}}{\bf \cdot \sigma }
_{k}/2}
\]
changing $(p,\widehat{{\bf d}})\rightarrow \{p\mu _{k},\widehat{{\bf d}}
_{k}\}_{k=1}^{N}$. So we have a Hamiltonian $\widehat{{\sf H}}_{p}^{{\bf %
Y\cdot S}}={\cal H}$ for 1-D closed chains, hence ${\cal H}$ is similar to $%
H_{xxx}^{tbc}$ (instead of $H_{xxz}^{tbc}$). The frustration condition and
the boundary's twisting $\Omega $ are, of course, the same of the case above.

It is important to note that we can construct ${\sf H}_{pq}^{{\bf S}}$ from $%
GL_{q}\left( 2\right) $ and $\widehat{{\cal G}}_{{\bf R}^{q}}$ $\supset
\widehat{{\cal G}}_{{\bf R}^{{\bf p}q}}$ (using $\Psi ^{\left( k\right)
}=p^{-\widehat{{\bf d}}{\bf \cdot \sigma }/2}$ for all $k$), but we want to
encode the Moriya's vectors content on a proper algebraic structure instead
on its internal symmetries.

Concluding, we have shown that the integrability of the DM interaction rest
on the underlying quantum group structure $GL_{pq}\left( 2\right) $,
connecting the 2-cocycle twisting $\Phi $ with the rotation that maps the
Hamiltonian in a $H_{xxz}$ like one. The existence of this map is well
understood in terms of the no frustration of the twisting along the whole
chain.

\subsection{Acknowledgments}

H. M. thank to CONICET and S.G. thank to CNEA, Argentina, for financial
support. The author also thank to C. Batista for useful discussions.


\end{document}